\documentclass[aps,10pt,prd,superscriptaddress,preprintnumbers,
nofootinbib,showpacs,twocolumn,preprintnumbers,tightenlines]{revtex4-2} 
\usepackage{xcolor}
\usepackage{slashed}
\usepackage{eucal} 
\usepackage{bm}
\usepackage{graphicx}
\usepackage[hypertexnames=false]{hyperref}
\usepackage{amsmath,amssymb,amsfonts,latexsym}
\begin{document}
\preprint{INHA-NTG-08/2026}
\author{June-Young Kim}
\email[ E-mail: ]{jun-young.kim@inha.ac.kr}
\affiliation{Department of Physics and Institute of Quantum Science,
  Inha University, Incheon 22212, Republic of Korea} 

  \author{Hyun-Chul Kim}
\email[ E-mail: ]{hchkim@inha.ac.kr}
\affiliation{Department of Physics and Institute of Quantum Science,
  Inha University, Incheon 22212, Republic of Korea} 

\title{Multipole structure of the $N \to \Delta$ Transition Generalized Parton Distributions}
\begin{abstract}
We establish the multipole structure of the $N \to \Delta$ transition
at the level of the generalized parton distributions (GPDs). 
We decompose the four transition GPDs into one monopole, two dipole,
and one quadrupole components in the transverse plane by a multipole
expansion of the covariant transition matrix element in terms of the
three-dimensional spin-transition tensors and the transverse momentum
transfer. These multipole components are in one-to-one correspondence
with the light-front helicity amplitudes. In the zero-skewness limit,
the multipole GPDs define impact-parameter transition densities, which
generalize the transverse transition charge densities to the
$x$-dependent level. These transition densities arise from
non-diagonal matrix elements between two distinct hadronic states and
must therefore be distinguished from the diagonal densities of the
nucleon and the $\Delta$. The multipole transition densities visualize  
the monopole, dipole, and quadrupole structures of the partonic $N \to
\Delta$ transition in the transverse plane.
\end{abstract}
\maketitle
%
%
%

\section{Introduction}
\label{sec:1}
Understanding the internal structure of the $\Delta(1232)$, the first excited
state of baryons, is a fundamental problem in hadron
physics~\cite{Pascalutsa:2006up}. Since the ephemeral $\Delta$ decays
strongly into $\pi N$~\cite{PDG:2026} and cannot be used as a target,
its structure is usually accessed through the electromagnetic (EM) $N
\to \Delta$ transition, characterized by the magnetic dipole ($M1$),
electric quadrupole ($E2$), and Coulomb quadrupole ($C2$) form
factors~\cite{Jones:1972ky}. Experiments~\cite{Beck:1997ew,
  Beck:1999ge, Blanpied:1997zz, Blanpied:2001ae, Mertz:1999hp,
  Sparveris:2004jn, Pospischil:2000ad, Frolov:1998pw, Joo:2001tw,
  Ungaro:2006df, Aznauryan:2009mx} based on pion photo- and
electroproduction off the nucleon have been carried out over decades,
and established that the $M1$ transition dominates, whereas the
measured quadrupole ratios $R_{EM}$ and $R_{SM}$ are only a few
percent in magnitude. These small but nonvanishing quadrupole
amplitudes indicate the deformation of the
$\Delta$~\cite{AlexandrouDeformation}, which arises from the pion
cloud in dynamical models~\cite{Kamalov:1999hs, Sato:2000jf} or from
the polarization of the vacuum (the Dirac continuum), as shown in the 
chiral quark-soliton model ($\chi$QSM)~\cite{Kim:2019gka}. In the
transverse charge densities constructed from the transition form
factors, these amplitudes appear as dipole and quadrupole patterns in
the transverse plane~\cite{Carlson:2007xd}.

The form factors, however, provide information only on the densities in the
transverse plane. A more complete description of the transition is
furnished by the generalized parton distributions (GPDs), 
which correlate the longitudinal momentum of the quark with its
transverse position inside the baryon~\cite{Muller:1994ses, Ji:1996ek,
Radyushkin:1997ki, Goeke:2001tz, Diehl:2003ny, Belitsky:2005qn}. At zero
skewness, the GPDs provide a probabilistic interpretation in
impact-parameter space or in the transverse plane~\cite{Burkardt:2000za,
  Burkardt:2002hr}. The concept of the GPDs can be extended to the
nucleon-to-resonance transition matrix elements, which makes it
possible to image the partonic structure of baryon resonances. Recently, hard
exclusive $\pi^{-}\Delta^{++}$ electroproduction was measured at the
Thomas Jefferson National Accelerator Facility
(JLab)~\cite{CLAS:2023akb}, exclusive $\pi\Delta(1232)$ production and
the deeply virtual Compton process $e N \to e \gamma \pi N$ were
analyzed theoretically~\cite{Kroll:2022roq,
  Semenov-Tian-Shansky:2023bsy}, and dedicated research programs on
transition GPDs will soon be available at the JLab upgrades and the
Electron-Ion Collider (EIC)~\cite{Diehl:2024bmd, AbdulKhalek:2021gbh}.

On the theoretical side, the vector $N \to \Delta$ transition GPDs
were first parametrized in Ref.~\cite{Goeke:2001tz} in terms of three
functions. A fourth Lorentz structure was subsequently proposed to
match the number of independent light-front helicity
amplitudes~\cite{Belitsky:2005qn}, but it turned out not to be
independent of the original three. A complete and independent fourth
structure was identified only recently in Ref.~\cite{Kim:2024hhd}, and
its kinematic and dynamical implications were explored in
Refs.~\cite{Kim:2024hhd, Kim:2025ilc, Kim:2025ves}. 

The aim of the present work is to establish the multipole structure of
the $N \to \Delta$ transition at the level of the GPDs. To this end,
we perform a multipole expansion of the covariant matrix element in
terms of the three-dimensional spin-transition tensors and the
transverse momentum transfer. The four transition GPDs are then
decomposed into one monopole, two dipole, and one quadrupole
components in the transverse plane, each of which corresponds to a
definite light-front helicity amplitude. The multipole GPDs extend the
multipole structure of the transition form factors~\cite{Jones:1972ky}
to the $x$-dependent level. In the zero-skewness limit, they define
impact-parameter transition densities, which generalize the transverse
charge densities of Ref.~\cite{Carlson:2007xd} and visualize the
partonic structure of the $N \to \Delta$ transition.

The present paper is organized as follows: Section~\ref{sec:2} reviews the
covariant parametrization of the $N \to \Delta$ transition GPDs and
their relation to the EM transition form factors. In Sec.~\ref{sec:3} we
introduce the spin-transition tensors in the symmetric light-front
frame, with which we derive the multipole expansion of the matrix element
and the corresponding light-front helicity amplitudes. In
Sec.~\ref{sec:4} we take the zero-skewness limit and visualize the
impact-parameter transition densities in the monopole, dipole, and
quadrupole channels. Section~\ref{sec:5} is devoted to the summary and
conclusions. Technical details are compiled in the Appendices.

\section{General formalism}
\label{sec:2}
In this section, we recapitulate the general formalism for the
$N \to \Delta$ transition GPDs. We define the nonlocal light-cone
matrix element, introduce the two covariant bases of the transition
GPDs, and relate their first Mellin moments to the EM transition form
factors, which serve as the starting point of the multipole analysis
in Sec.~\ref{sec:3}.

\subsection{Generalized parton distributions}
The transition matrix element of the nonlocal vector
operator between the proton $N^{+}$ and the $\Delta^+$ baryon states
is defined as 
\begin{align}
&\mathcal{M}^{[\slashed{n}]}_{\mathrm{GPDs}} \nonumber \\[0.5ex]
&=
\int \frac{d\lambda}{2\pi} e^{i\lambda x}
\langle \Delta^{+} |
\bar{\psi}(-\lambda n/2) \slashed{n} \tau^{3}
\psi(\lambda n/2)
| N^{+} \rangle ,
\label{eq:mat_GPD}
\end{align}
where $\psi$ denotes the quark field operator, $\lambda$ is the
light-cone separation between the two fields, and $n$ is a lightlike
vector satisfying $n^{2}=0$. We use the shorthand $|N^{+}\rangle
\equiv |N^{+}(p,\sigma)\rangle$ and $\langle \Delta^{+}| \equiv
\langle \Delta^{+}(p',\sigma')|$. The four-momenta and spin
polarizations of the initial and final states are denoted by 
$(p,\sigma)$ and $(p',\sigma')$, respectively. These arguments are
suppressed hereafter. The matrix $\tau^{3}$ denotes the SU(2) flavor
matrix for the $N\to\Delta$ transition.

This matrix element is parametrized in terms of the transition GPDs.
Three vector transition GPDs were first introduced in
Ref.~\cite{Goeke:2001tz}, in analogy with the EM $N\to\Delta$
transition form factors, to which their first Mellin moments are
related. As discussed in Sec.~\ref{sec:1}, however, four independent
Lorentz structures are required to match the number of light-front
helicity amplitudes~\cite{Belitsky:2005qn}, and the complete
parametrization was established in Ref.~\cite{Kim:2024hhd}.

The four $N^+\to\Delta^+$ transition GPDs are defined through the 
light-cone vector-current matrix element,
\begin{align}
\mathcal{M}^{[\slashed{n}]}_{\mathrm{GPDs}}
&= \sqrt{\frac{2}{3}}\,
\bar{u}_{\alpha}
\left[
\sum_{I=1,2,3,X}
G_I(x,\xi,t)\,
\mathcal{K}_{I}^{\alpha\mu} n_{\mu}
\right]
u .
\label{eq:Belitsky_para}
\end{align}
Here $G_I(x,\xi,t)$, with $I=1,2,3,X$, denote the four
$N^+\to\Delta^+$ transition GPDs, and $\sqrt{2/3}$ is the isospin
Clebsch--Gordan coefficient for this transition. We use the symmetric
kinematics
\begin{align}
    P=\frac{p+p'}{2}, \qquad
    \Delta=p'-p ,
\end{align}
and define
\begin{align}
    t=\Delta^2, \qquad
    \xi=-\frac{\Delta\cdot n}{2P\cdot n}.
    \label{eq:skewness}
\end{align}
The variable $x$ denotes the average light-cone momentum fraction
carried by the quark with respect to $P\cdot n$, while $\xi$ is the
skewness parameter measuring the longitudinal momentum transfer and
$t$ is the invariant momentum transfer squared. The spinors 
$\bar{u}_{\alpha}\equiv \bar{u}_{\alpha}(p',\sigma')$ and
$u \equiv u(p,\sigma)$ denote the Rarita--Schwinger spinor
for the outgoing $\Delta$ baryon and the Dirac spinor for the incoming
nucleon, respectively. They satisfy the on-shell Dirac equations 
\begin{align}
    \bar{u}_{\alpha} \slashed{p}'
    &= m_{\Delta}\,\bar{u}_{\alpha} ,
    &
    \slashed{p}\,u
    &= m_N\,u,
\end{align}
with $p'^2=m_\Delta^2$ and $p^2=m_N^2$. These on-shell conditions
yield 
\begin{subequations}
\label{eq:on_shell}
\begin{align}
\Delta \cdot P &= \frac{m^{2}_{\Delta}-m^{2}_{N}}{2}, \\
P^{2}+\frac{\Delta^{2}}{4}
&= \frac{m^{2}_{N}+m^{2}_{\Delta}}{2}.
\end{align}
\end{subequations}
The Rarita--Schwinger spinor
also satisfies the subsidiary condition
\begin{align}
    \bar{u}_{\alpha} \gamma^\alpha = 0 .
\end{align}
The tensors $\mathcal{K}_{I}^{\alpha\mu}$ encode the independent
Lorentz structures constructed from the available four-vectors $p$ and
$p'$, the metric tensor $g^{\mu\nu}$, the Levi-Civita tensor
$\epsilon^{\mu\nu\rho\sigma}$, and the Dirac matrices. Their explicit
expressions are given 
by~\cite{Kim:2024hhd}
\begin{subequations}
\label{eq:2}
\begin{align}
\mathcal{K}^{\alpha \mu}_{1}
&= \frac{1}{m_{N}}
\left( \Delta^{\alpha} \gamma^{\mu}
- \slashed{\Delta} g^{\alpha \mu} \right)\gamma_{5}, \\
\mathcal{K}^{\alpha \mu}_{2}
&= \frac{2}{m^{2}_{N}}
\left( \Delta^{\alpha} P^{\mu}
- (\Delta \cdot P) g^{\alpha \mu} \right)\gamma_{5}, \\
\mathcal{K}^{\alpha \mu}_{3}
&= \frac{1}{m^{2}_{N}}
\left( \Delta^{\alpha} \Delta^{\mu}
- \Delta^{2} g^{\alpha \mu} \right)\gamma_{5}, \\[1ex]
\mathcal{K}^{\alpha \mu}_{X}
&= m_{N} n^{\alpha} \gamma^{\mu} \gamma_{5}.
\end{align}
\end{subequations}
The first three structures correspond to the Belitsky--Radyushkin (BR) 
parametrization, whereas the last one was introduced in
Ref.~\cite{Kim:2024hhd}. The choice of basis, however, is not unique.
Another commonly used basis is that of the
Goeke--Polyakov--Vanderhaeghen~(GPV) GPDs~\cite{Goeke:2001tz},
motivated by the Jones--Scadron multipole form
factors~\cite{Jones:1972ky}. The GPV GPDs are related to the BR GPDs 
by the linear transformation
\begin{widetext}
\begin{align}
&3m_{\Delta}m_{N}(m_{\Delta}+m_{N})
\left(
\begin{array}{c}
H_{M} \\
H_{E} \\
H_{C}
\end{array}
\right)
=
\left(
\begin{array}{ccc}
(3m_{\Delta}+m_{N})(m_{\Delta}+m_{N})-t
& 2(m_{\Delta}^{2}-m_{N}^{2})
& 2t \\
m_{\Delta}^{2}-m_{N}^{2}+t
& 2(m_{\Delta}^{2}-m_{N}^{2})
& 2t \\
4m_{\Delta}^{2}
& 2(3m_{\Delta}^{2}+m_{N}^{2}-t)
& 2(m_{\Delta}^{2}-m_{N}^{2}+t)
\end{array}
\right)
\left(
\begin{array}{c}
m_{N}G_{1} \\
m_{\Delta}G_{2} \\
m_{\Delta}G_{3}
\end{array}
\right).
\label{eq:connect}
\end{align}
\end{widetext}
We use Eq.~\eqref{eq:connect} as the working definition for converting 
between the GPV and BR bases. The full covariant decomposition of the
GPV parametrization will not be reproduced here. Its explicit form is
given in Refs.~\cite{Goeke:2001tz, Kim:2024hhd}.

\subsection{Electromagnetic form factors}
The first moment of the nonlocal operator in Eq.~\eqref{eq:mat_GPD} 
reduces to the local EM current and thus relates the transition GPDs
in Eq.~\eqref{eq:Belitsky_para} to the corresponding EM transition
form factors. Note that only the isovector component of the EM current
contributes to the $N \to \Delta$ transition. In the Jones--Scadron
covariant parametrization~\cite{Jones:1972ky}, written with the same
Lorentz structures as in Eq.~\eqref{eq:2}, the EM current matrix
element is expressed as
\begin{align}
\mathcal{M}^{[\gamma^\mu]}_{\mathrm{FF}}
&= \langle \Delta^{+} |
\bar{\psi}(0) \gamma^{\mu} \frac{\tau^{3}}{2} \psi(0)
| N^{+} \rangle \nonumber \\
&= \sqrt{\frac{2}{3}}\,
\bar{u}_{\alpha}
\left[
\sum_{I=1,2,3}
G^{*}_{I}(t)\,
\mathcal{K}_{I}^{\alpha\mu}
\right]
u .
\end{align}
Here the Lorentz structures are identical to those in
Eq.~\eqref{eq:2}, but the dynamical quantities are form factors, 
depending only on the invariant momentum transfer squared $t$. The
asterisk is used to distinguish the EM form factors from the GPDs.
The first moments of the transition GPDs are then given by
\begin{subequations}
\begin{align}
\int dx \, G_{1,2,3}(x,\xi,t) &= 2G^{*}_{1,2,3}(t), \\[0.5ex]
\int dx \, G_{X}(x,\xi,t) &= 0.
\label{eq:GX_sumrule}
\end{align}
\end{subequations}
Similarly, the first moments of the GPV GPDs yield the multipole form
factors
\begin{align}
\int dx \, H_{M,E,C}(x,\xi,t) &= 2G^{*}_{M,E,C}(t),
\end{align}
where $G^{*}_{M}(t)$, $G^{*}_{E}(t)$, and $G^{*}_{C}(t)$ denote the
magnetic dipole, electric quadrupole, and Coulomb quadrupole form
factors, respectively. They are related to $G^{*}_{1,2,3}(t)$ by the
same linear transformation as in Eq.~\eqref{eq:connect}.

\section{Multipole expansion}
\label{sec:3}
The Lorentz covariant structure of the matrix element in
Eqs.~\eqref{eq:Belitsky_para} and~\eqref{eq:2} can be analyzed by
performing a multipole expansion in terms of the three-dimensional
polarization tensors and the angular dependence of the transverse
momentum transfer. This
expansion provides a natural way to relate the angular structure of
the matrix element to the underlying spatial distributions. We first
introduce the polarization tensors, specify the symmetric-frame
kinematics, and then construct the corresponding multipole basis and
helicity amplitudes. 

\subsection{Polarization tensors in spherical basis}
The spin structure of the diagonal matrix element can be expressed in
terms of the polarization tensor, which is well understood (see
Ref.~\cite{Varshalovich:1988ifq}). We extend this construction to
the non-diagonal matrix element in Eq.~\eqref{eq:mat_GPD} by choosing a
normalization consistent with the diagonal case: 
\begin{align}
T_{LM}(\sigma',\sigma) \equiv \frac{\sqrt{2L+1}}{\sqrt{(2S+1)}} C^{S'
  \sigma'}_{S\sigma L M},
\label{eq:pt_sb}
\end{align}
where $C^{S' \sigma'}_{S\sigma L M}$ denotes Clebsch--Gordan~(CG)
coefficients. For the $N\to\Delta$ transition, the spin quantum
numbers of the initial and final states are fixed to $S=1/2$ and
$S'=3/2$, respectively. The allowed polarization tensors $T_{LM}
\equiv T_{LM}(\sigma',\sigma)$ are then restricted by the
angular-momentum selection rule to 
\begin{subequations}
\label{eq:PT_s}
\begin{align}
&T_{1M}, && M=\pm 1, \label{eq:PT_sa} \\[1ex]
&T_{2M}, && M=0,\pm1,\pm2 \label{eq:PT_sb} .
\end{align}
\end{subequations}
Thus, only the dipole $(L=1)$ and quadrupole $(L=2)$ transitions
contribute, while the monopole $(L=0)$ and higher multipoles
$(L\geq3)$ are excluded by this constraint. 

\subsection{Polarization tensors in Cartesian basis}
The spherical polarization tensors in Eq.~\eqref{eq:PT_s} can be
represented in terms of Cartesian spin-transition tensors. We
introduce the corresponding Cartesian vector and rank-two tensor
components as 
\begin{align}
V^{i} \equiv V^{i}(\sigma',\sigma), \qquad
Q^{ij} \equiv Q^{ij}(\sigma',\sigma),
\label{eq:PT_c}
\end{align}
with $i,j=1,2,3$. Using the standard spherical--Cartesian
convention~\cite{Varshalovich:1988ifq}, the vector components are
related by 
\begin{align}
T_{10} = V^{3}, \qquad
T_{1\pm1} = \mp \frac{1}{\sqrt{2}} (V^{1} \pm i V^{2}),
\label{eq:vr}
\end{align}
and the rank-two components by
\begin{subequations}
\begin{align}
T_{20} &= \sqrt{\frac{3}{2}}\, Q^{33}, \\[0.5ex]
T_{2\pm1} &= \mp (Q^{13} \pm i Q^{23}), \\[0.2ex]
T_{2\pm2} &= \frac{1}{2}(Q^{11}-Q^{22} \pm 2i Q^{12}) .
\end{align}
\end{subequations}
Here $Q^{ij}$ is symmetric and traceless,
\begin{align}
Q^{ij}=Q^{ji}, \qquad Q^{ii}=0 .
\end{align}

These spin-transition tensors can be represented by coupling the
spin-1 polarization vector to the Pauli spin operator. We define 
\begin{subequations} 
\label{eq:CT} 
\begin{align} 
V^{i}& = \sqrt{\frac{3}{2}} \sum_{\lambda'} C^{\frac32 \sigma' }_{1
       \lambda' \frac12 \sigma}
       \hat{\epsilon}^{*i}_{\lambda'}, \label{eq:CTa} \\  
Q^{ij}& = \sum_{\lambda' s'} C^{\frac32 \sigma' }_{1 \lambda' \frac12
        s'} \frac{1}{2}\sigma^{ \{i}_{s' \sigma} \hat{\epsilon}^{*j\}
        }_{\lambda'}. \label{eq:CTb}  
\end{align} 
\end{subequations}
where $a^{ \{i} b^{j\} } = a^{i}b^{j}+a^{j}b^{i}$.
The vector $\hat{\epsilon}^{*i}_{\lambda'}$ is the complex conjugate
of the spherical polarization vector, with 
\begin{align}
\hat{\epsilon}^{i}_{0}= (0,0,1), \quad \hat{\epsilon}^{i}_{\pm 1}= \mp
  \frac{1}{\sqrt{2}}(1,\pm i,0). 
\end{align}
It is related to the Cartesian-to-spherical transformation matrix
$U^{\mathrm{CS}}$ through 
\begin{align}
\hat{\epsilon}^{*i}_{\lambda'}= (-1)^{-\lambda'}
  U^{\mathrm{CS}}_{i,-\lambda'}. 
\label{eq:CS}
\end{align}
The quantity $(\sigma^i)_{s'\sigma}$ denotes the Pauli matrix
element. 

Equation~\eqref{eq:CTa} is therefore directly related to
$T_{1M}$ in Eq.~\eqref{eq:PT_sa} by the spherical--Cartesian
transformation~\eqref{eq:CS}. In the rank-two case, 
Eq.~\eqref{eq:CTb}, the Pauli matrix element is first rewritten as a 
rank-one spin tensor, which introduces the CG coefficient
for coupling the spin-$1/2$ state to a rank-one spin operator:
\begin{align}
(\sigma^i)_{s'\sigma}  = \sqrt{3} \, \sum_{\mu=-1,0,1} C^{\frac12
  s'}_{\frac12 \sigma 1 \mu} \hat{\epsilon}^{* i}_{\mu}. 
\label{eq:a1}
\end{align}
The symmetric-traceless projection of the two spin-1 vectors in 
Eq.~\eqref{eq:CTb} then selects the $L=2$ component and introduces the 
CG coefficient for $1\otimes 1\to 2$. Together with the
CG coefficient already present in Eq.~\eqref{eq:CTb}, these
coefficients are recoupled into the single coefficient that defines
$T_{2M}$ in Eq.~\eqref{eq:PT_sb}. In contrast, the antisymmetric part of the
same product projects onto the $L=1$ component. Using
Eq.~\eqref{eq:a1}, one readily obtains 
\begin{align}
i \epsilon^{ijk}  V^{k}
&=
\sqrt{\frac{3}{2}} \, \sum_{\lambda' s'}
C^{\frac32 \sigma'}_{1 \lambda' \frac12 s'}
\left(
\sigma^{i}_{s'\sigma}\hat{\epsilon}^{*j}_{\lambda'}
-
\sigma^{j}_{s'\sigma}\hat{\epsilon}^{*i}_{\lambda'}
\right).
\label{eq:CTc}
\end{align}
This relation reflects the fact that the antisymmetric product of two 
spin-1 vector operators transforms as a spin-1 tensor. 

Equations~\eqref{eq:CT}--\eqref{eq:CTc} provide the Cartesian working 
representation of the spin-transition tensors. This representation will
serve as the practical working expression for the explicit evaluation of
the Rarita--Schwinger and Dirac spinors in the multipole expansion
(see Appendices~\ref{app:a}--\ref{app:c}).

\subsection{Symmetric frame in light-front quantization
  \label{sec:SF}}
We specify the light-front (LF) kinematics for the non-diagonal
$N\to\Delta$ matrix element entering the GPD parametrization in
Eqs.~\eqref{eq:Belitsky_para} and \eqref{eq:2}. The LF components of a
four-vector are defined as $a=(a^+,a^-,\bm{a}_\perp)$, with
$a^\pm=(a^0\pm a^3)/\sqrt{2}$. The scalar product is then given by 
\begin{align}
a\cdot b = a^+b^- + a^-b^+ - \bm{a}_\perp\cdot\bm{b}_\perp .
\end{align}
We work in the standard symmetric frame commonly used for GPDs,
\begin{subequations}
\label{eq:sy_frame}
\begin{align}
P&=(P^+,P^-,\bm{0}_\perp), \label{eq:sy_frame_a}\\[1ex] 
\Delta&=(\Delta^+,\Delta^-,\bm{\Delta}_\perp). \label{eq:sy_frame_b}
\end{align}
\end{subequations}
The plus component of the momentum transfer in
Eq.~\eqref{eq:sy_frame_b} is determined to be $\Delta^+=-2\xi P^+$
from Eq.~\eqref{eq:skewness} by choosing the light-cone vector as 
$n^\mu=(1,0,0,-1)/(\sqrt{2}P^+)$ in Minkowski coordinates. The
on-shell conditions in Eq.~\eqref{eq:on_shell}, together with the
symmetric-frame condition in Eq.~\eqref{eq:sy_frame}, fix the minus
components of Eq.~\eqref{eq:sy_frame} as 
\begin{subequations}
\label{eq:GPD_frame1}
\begin{align}
P^-
&=
\frac{
2\xi(m_\Delta^2-m_N^2)+2m_\Delta^2+2m_N^2+\bm{\Delta}_\perp^2
}
{8P^+(1-\xi^2)}, \\
\Delta^-
&=
\frac{
2(m_\Delta^2-m_N^2)+\xi(2m_\Delta^2+2m_N^2+\bm{\Delta}_\perp^2)
}
{4P^+(1-\xi^2)} .
\end{align}
\end{subequations}
In this frame, the remaining angular dependence enters only through
the direction of the transverse momentum transfer, 
\begin{align}
\bm{\Delta}_\perp = |\bm{\Delta}_\perp| \, (\cos\theta, \sin\theta).
\label{eq:AMA}
\end{align}
It can be organized in terms of the irreducible two-dimensional tensors
\begin{subequations}
\label{eq:MT_basis}
\begin{align}
&X_{0} \hspace{0.45cm}=1,  &&(M=0), \\[1.4ex]
&X^{i}_{1}(\theta)=\frac{\bm{\Delta}^{i}_{\perp}}{|\bm{\Delta}_{\perp}|},
                           &&(M=\pm 1), \\ 
&X^{ij}_{2}(\theta)=\frac{\bm{\Delta}^{i}_{\perp}
  \bm{\Delta}^{j}_{\perp}}{|\bm{\Delta}_{\perp}|^{2}}-\frac{1}{2}\delta^{ij},  
                           &&(M=\pm 2), 
\end{align}
\end{subequations}
which form a convenient tensor basis for the multipole expansion.
 
In the forward limit $\xi,t\to 0$, the
four-momenta~\eqref{eq:sy_frame} reduce to 
\begin{subequations}
\label{eq:GPD_frame2}
\begin{align}
P &= \left(P^+, \frac{m_\Delta^2+m_N^2}{4P^+}, \bm{0}_\perp \right), \\
\Delta &= \left(0, \frac{m_\Delta^2-m_N^2}{2P^+}, \bm{0}_\perp \right).
\end{align}
\end{subequations}
For $m_\Delta\neq m_N$, the minus component of the momentum transfer,
$\Delta^{-}$, remains nonzero, so that $\Delta^\mu\neq0$. The
vanishing momentum transfer is recovered only in the diagonal case,
$m_\Delta=m_N$. 

\subsection{Matrix element in multipole basis}
In the symmetric frame specified in Sec.~\ref{sec:SF}, evaluating the
Dirac and Rarita--Schwinger spinors allows the covariant
parametrizations in Eqs.~\eqref{eq:Belitsky_para} and~\eqref{eq:2} to
be mapped onto the four independent multipole structures defined in
the bases of Eqs.~\eqref{eq:PT_c} and~\eqref{eq:MT_basis},
\begin{align}
&\mathcal{M}^{[\slashed{n}]}_{\mathrm{GPDs}}
=
Q^{33} X_{0}\, M_{0}
- i \epsilon^{3kl} V^{l} X^{k}_{1}(\theta)
\frac{|\bm{\Delta}_{\perp}|}{m_{N}}\, M^{V}_{1}
\nonumber \\[1ex]
&\quad
+ Q^{3k} X^{k}_{1}(\theta)
\frac{|\bm{\Delta}_{\perp}|}{m_{N}}\, M^{Q}_{1}
+ Q^{kl} X^{kl}_{2}(\theta)
\frac{|\bm{\Delta}_{\perp}|^{2}}{m^{2}_{N}}\, M_{2}.
\label{eq:multipole_slashn}
\end{align}
The coefficients $M_{0}$, $M^{V}_{1}$, $M^{Q}_{1}$, and $M_{2}$ are
scalar functions of $(x,\xi,t)$, which define the corresponding
multipole GPDs. Their subscripts indicate the rank of the transverse
multipole tensor: $M_{0}$ is associated with the scalar component
$X_{0}$, $M^{V}_{1}$ and $M^{Q}_{1}$ with the two independent vector
components proportional to $X_{1}^{k}$, and $M_{2}$ with the rank-two
tensor $X_{2}^{kl}$. These functions are related to the
covariant GPDs $G_{1,2,3,X}$ through linear combinations, given
explicitly in Appendix~\ref{app:d}. 

\subsection{Light-front helicity amplitudes}
The multipole decomposition can also be expressed in terms of
light-front helicity amplitudes. This representation is useful because
the transverse multipole rank is reflected in the corresponding
helicity-flip structure. We define the helicity amplitudes by
factoring out the azimuthal phase, 
\begin{align}
\mathcal{M}^{[\slashed{n}]}_{\mathrm{GPDs}} =
2 e^{i(\sigma-\sigma')\theta}
A_{\sigma'\sigma},
\end{align}
where $\theta$ is the azimuthal angle of $\bm{\Delta}_{\perp}$, as
defined in Eq.~\eqref{eq:AMA}. The light-front helicity amplitudes 
$A_{\sigma'\sigma} \equiv A_{\sigma'\sigma}(x,\xi,t)$ then depend only
on the GPD variables.

Using the multipole representation in Eq.~\eqref{eq:multipole_slashn},
one obtains
\begin{subequations}
\begin{align}
A_{\frac12 \frac12}
&= -A_{-\frac12 -\frac12}
= \dfrac{M_{0}}{\sqrt{6}},  \\
A_{\frac32 \frac12}
&= A_{-\frac32 -\frac12}
= \dfrac{|\bm{\Delta}_{\perp}|}{8m_{N}}
\left(2\sqrt{3}M^{V}_{1}-\sqrt{2}M^{Q}_{1}\right), \\
A_{\frac12 -\frac12}
&= A_{-\frac12 \frac12}
= \dfrac{|\bm{\Delta}_{\perp}|}{8m_{N}}
\left(2M^{V}_{1}+\sqrt{6}M^{Q}_{1}\right),\\
A_{\frac32 -\frac12}
&= -A_{-\frac32 \frac12}
= -\dfrac{|\bm{\Delta}_{\perp}|^{2}}{2\sqrt{2}m^{2}_{N}} M_{2}.
\end{align}
\end{subequations}
The power of $|\bm{\Delta}_{\perp}|$ in each term matches the helicity
difference $|\sigma'-\sigma|$. Accordingly, the helicity-conserving
amplitude is governed by the monopole GPD $M_{0}$, the amplitudes with
$|\sigma'-\sigma|=1$ by the two independent dipole GPDs $M^{V}_{1}$
and $M^{Q}_{1}$, and the amplitude with $|\sigma'-\sigma|=2$ by the
quadrupole GPD $M_{2}$. The light-front helicity amplitudes therefore
provide a direct physical interpretation of the multipole GPDs
introduced above. 

\section{Mechanical interpretation of
  the transition matrix elements} 
\label{sec:4}
In this section, we turn to the spatial interpretation of the
multipole GPDs derived in Sec.~\ref{sec:3}. In the zero-skewness
limit, their two-dimensional Fourier transforms define the
impact-parameter transition densities, which visualize the monopole,
dipole, and quadrupole structures of the $N \to \Delta$ transition in
the transverse plane. We also discuss how these transition densities
should be interpreted in comparison with the diagonal densities of the
nucleon and the $\Delta$.

\subsection{Impact-parameter distributions}
We now consider the zero-skewness limit, $\xi \to 0$. In this limit,
$\Delta^{+}=0$, so that the initial and final hadron states carry the
same plus momentum. The Fourier transform with respect to
$\bm{\Delta}_{\perp}$ can therefore be interpreted as resolving the
transverse structure of the matrix element at fixed longitudinal
momentum, rather than as an overlap between states with different plus
momenta~\cite{Burkardt:2000za,Burkardt:2002hr}. In this limit, 
\begin{align}
|\bm{\Delta}_{\perp}|^{2}=-t .
\end{align}
We define the corresponding impact-parameter-space distribution for 
the $N\to\Delta$ transition by 
\begin{align}
\frac{1}{2}
\int \frac{d^{2}\Delta_{\perp}}{(2\pi)^{2}}\,
e^{-i \bm{\Delta}_{\perp}\cdot\bm{b}}
\mathcal{M}^{[\slashed{n}]}_{\mathrm{GPDs}}  \underset{\xi \to 0}{=}
  \rho(x,\bm{b};\sigma',\sigma), 
\label{eq:rho_def}
\end{align}
where $\bm{b}$ denotes the transverse impact parameter:
\begin{align}
\bm{b} = |\bm{b}| \, (\cos\phi, \sin\phi).
\label{eq:IPD_b}
\end{align}

Since the matrix element is nondiagonal in the hadronic states,
$\rho(x,\bm{b};\sigma',\sigma)$ is not a probability density of quarks
inside either the nucleon or the $\Delta$. Rather, it is an
impact-parameter-space transition density associated with the
light-front quark operator in Eq.~\eqref{eq:mat_GPD}. At fixed
longitudinal momentum fraction $x$, it gives the dependence of the
$N\to\Delta$ transition amplitude on the transverse impact parameter
$\bm b$, defined with respect to the light-front transverse center of
momentum~\cite{Burkardt:2002hr}. The baryon mass difference does not
affect the definition of the transverse center of momentum as long as
$\Delta^{+}=0$, so that the impact-parameter representation is as well
defined as in the equal-mass case~\cite{Kim:2025ilc}. 

This construction may be regarded as an $x$-dependent extension of
transverse $N\to\Delta$ transition densities obtained at the
form-factor level~\cite{Carlson:2007xd}. A three-dimensional
representation in $(x,b_x,b_y)$ should therefore be read as showing
the $x$ and transverse impact-parameter dependence of the transition
density, not as an ordinary probability density. 

Applying the Fourier transform in Eq.~\eqref{eq:rho_def} to the
momentum-space multipole decomposition in
Eq.~\eqref{eq:multipole_slashn} gives the corresponding
impact-parameter-space representation. The transition density can then
be written in the same spin--multipole basis as 
\begin{align}
\rho(x,\bm{b}&;\sigma',\sigma)
= Q^{33}\rho_{0}(b) + \epsilon^{3kl}
               V^{l}X^{k}_{1}(\phi)\,\rho^{V}_{1}(b) \nonumber \\[1ex] 
&\quad
+i Q^{3k}X^{k}_{1}(\phi) \, \rho^{Q}_{1}(b) + Q^{kl}X^{kl}_{2}(\phi)
                                                                          \,\rho_{2}(b), 
\label{eq:rho_multipole}
\end{align}
where $b=|\bm{b}|$. The radial functions $\rho_0$, $\rho^{V}_1$,
$\rho^{Q}_1$, and $\rho_2$ are the Fourier--Bessel transforms of the
corresponding multipole GPDs $M_0$, $M^{V}_1$, $M^{Q}_1$, and
$M_2$. They represent the monopole, the two independent dipole, and
the quadrupole components of the $N\to\Delta$ transition density in
the transverse plane. The radial functions in
Eq.~\eqref{eq:rho_multipole} are defined as 
\begin{subequations}
\label{eq:rho_radial}
\begin{align}
\rho_{0}(x,b)
&=
\tilde{M}_{0}(x,0,b),
\\[1.5ex]
\rho^{V,Q}_{1}(x,b)
&=
\frac{1}{m_{N}}\frac{d}{db}\tilde{M}^{V,Q}_{1}(x,0,b),
\\[1ex]
\rho_{2}(x,b)
&=
-\frac{1}{m_{N}^{2}}\,
b\frac{d}{db}\frac{1}{b}\frac{d}{db}
\tilde{M}_{2}(x,0,b),
\end{align}
\end{subequations}
where the functions $\tilde{M}$ are defined as the Fourier transforms
of the corresponding multipole GPDs. 
\begin{align}
\tilde{M}(x,0,b) = \frac{1}{2} \int \frac{d^{2}\Delta_{\perp}}{(2\pi)^{2}}\,
e^{-i \bm{\Delta}_{\perp}\cdot\bm{b}}\,
M(x,0,t).
\end{align}

\subsection{Visualization of the partonic structure}
We now evaluate the transition density for specific polarization
states of the initial nucleon and final $\Delta$ baryon. For the
unpolarized quark operator~\eqref{eq:mat_GPD}, the transition density
between longitudinally polarized states ($\sigma'=\sigma=1/2$) is
given by 
\begin{align}
\rho (x,\bm{b};\sigma'=1/2,\sigma=1/2)
&=\sqrt{\frac{2}{3}} \, \rho_{0}(x,b).
\label{eq:LD}
\end{align}
Thus, only the monopole component contributes in this case.

For transversely polarized states, the transition density acquires
additional angular structures in the transverse plane. We choose the
transverse spin along the $x$ direction and define
\begin{subequations}
\begin{align}
|N, s_{x}=1/2  \rangle &=  \frac{1}{\sqrt{2}}\bigg{(}| \sigma=1/2
                         \rangle  + | \sigma=-1/2 \rangle \bigg{)},
  \\ 
|\Delta, s_{x}=1/2  \rangle &=  \frac{1}{\sqrt{8}}\bigg{(} \sqrt{3}|
                              \sigma=3/2 \rangle  + | \sigma=1/2
                              \rangle  \cr 
&- | \sigma=-1/2 \rangle - \sqrt{3}|  \sigma=-3/2 \rangle \bigg{)}, 
\end{align}
\end{subequations}
For these polarization states, one obtains
\begin{align}
&\rho(x,\bm{b};s'_x=1/2,s_x=1/2)=
\sqrt{\frac{1}{6}} \, \rho_{0}(x,b)  \cr
&+\rho^{V}_{1}(x,b) \sin\phi -\frac{\sqrt{6}}{4} \, \rho_{2}(x,b) \cos 2\phi.
\label{eq:TD}
\end{align}
The first term is the monopole contribution, while the terms
proportional to $\sin\phi$ and $\cos 2\phi$ represent the dipole and
quadrupole distortions of the transverse transition density,
respectively. 

\begin{figure}[htp]
\includegraphics[scale=0.7]{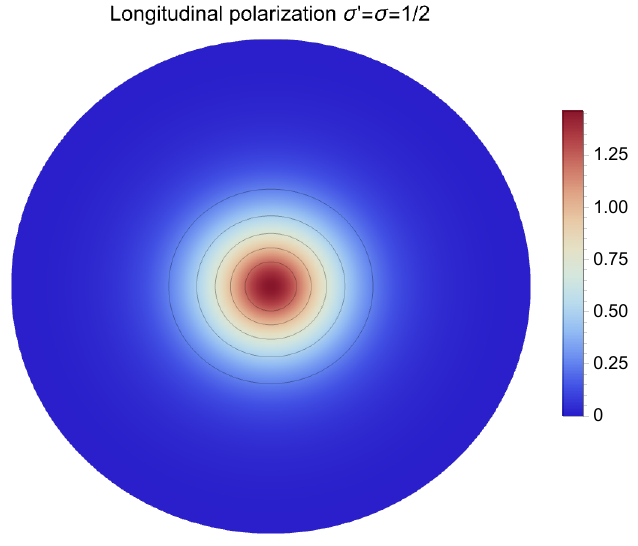}
\includegraphics[scale=0.7]{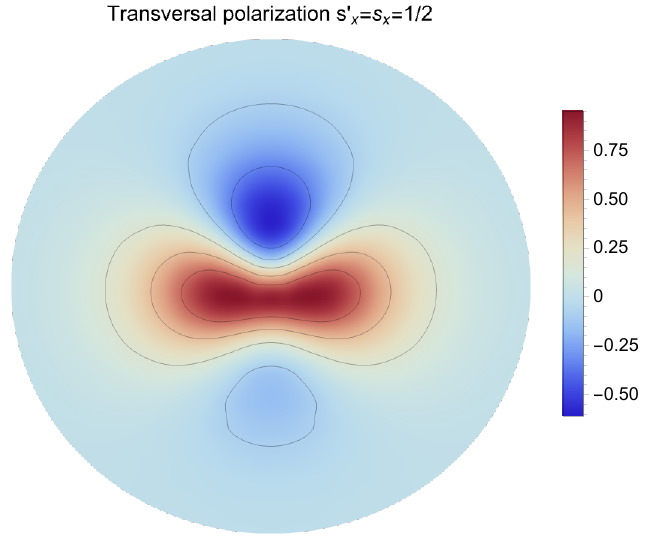}
\caption{Visualization of the transition densities when the nucleon
  and the $\Delta$ baryon are longitudinally (upper panel) and
  transversely (lower panel) polarized.} 
\label{fig:1}
\end{figure}

\begin{figure}[htp]
\includegraphics[scale=0.7]{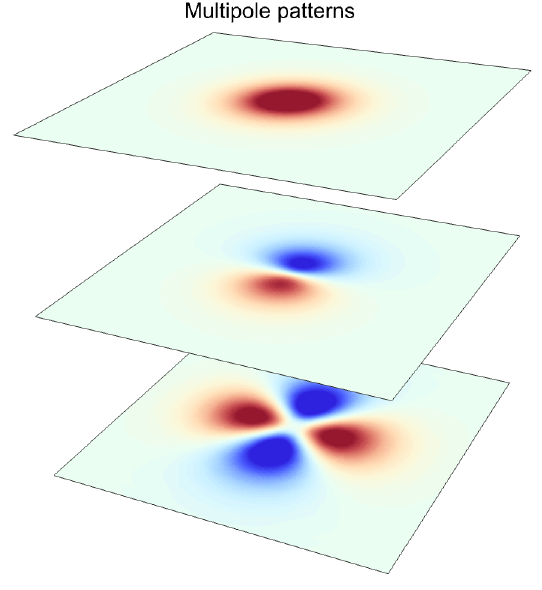}
\caption{Impact-parameter distributions of unpolarized quarks for the
  proton and $\Delta$ baryon transversely polarized along the $x$
  axis. The distributions exhibit richer multipole patterns,
  consisting of monopole (upper layer), dipole (middle layer), and
  quadrupole (lower layer) structures.} 
\label{fig:2}
\end{figure}

Using a multipole ansatz, we visualize Eqs.~\eqref{eq:LD} and
\eqref{eq:TD} in Fig.~\ref{fig:1}.  This ansatz is intended to
highlight the multipole patterns of the distributions rather than
their absolute strengths, which depend on the underlying dynamics and
are beyond the scope of the present study. The individual multipole
contributions are shown separately in Fig.~\ref{fig:2}. 

\subsection{Interpretation and discussion}
The physical meaning of the transition density must be carefully
distinguished from that of the diagonal densities of the nucleon and
the $\Delta$. The diagonal density, obtained from the $N \to N$ or
$\Delta \to \Delta$ matrix element at zero skewness, is positive
definite in its helicity-conserving component and answers the question
of where the quarks are located inside the given baryon. In contrast,
the transition density arises from the overlap of two distinct
hadronic states, so that it obeys no positivity condition and no
number sum rule. Its moments are instead given by the transition form 
factors, and the structure that survives in the forward limit obeys
the zero sum rule in Eq.~\eqref{eq:GX_sumrule} imposed by current
conservation~\cite{Kim:2025ilc}. It answers a different question,
namely, at which transverse
position the quark operator converts the nucleon into the $\Delta$.
For the same reason, the quadrupole pattern of the transition density
should not be identified with the intrinsic deformation of either the
nucleon or the $\Delta$ alone. It characterizes the spin-angular
structure of the $N \to \Delta$ overlap, whereas the intrinsic
deformation of each baryon is encoded in the corresponding diagonal
densities. The density interpretation of the forward-limit
($\bm{\Delta}_{\perp}=0$) transition matrix element was clarified in
Ref.~\cite{Kim:2025ilc}, where the partonic operator acts as a number
density in the partonic states while the matrix element connects
different hadron states. In the forward limit, the monopole GPD
$M_{0}$ becomes proportional to $G_{X}$ and thus to the
tensor-polarized transition parton density, whose magnitude was
estimated in the mean-field approach and found to be parametrically
small~\cite{Kim:2025ilc, Kim:2025ves}. The present work extends this
distinction to
the impact-parameter densities at nonzero transverse momentum
transfer, where the dipole and quadrupole structures appear.

A region with a large magnitude corresponds to a transverse position
where the quark operator gives a strong contribution to the transition
amplitude. Being an amplitude rather than a probability density, the
transition density carries a sign that reflects the relative phase of
the overlap. Consequently, different regions in the transverse plane
may interfere constructively or destructively in the integrated form
factor. 

The multipole structures encode the spin-angular content of the
transition, as visualized in Figs.~\ref{fig:1} and~\ref{fig:2}. When
both baryons are longitudinally polarized, only the monopole component 
contributes, as shown in Eq.~\eqref{eq:LD}, so that the transition
density is axially symmetric in the transverse plane (upper panel of
Fig.~\ref{fig:1}). For transverse polarization along the $x$ axis, the
azimuthal symmetry is broken according to Eq.~\eqref{eq:TD}. The
dipole term, proportional to $\sin\phi$, displaces the density
perpendicular to the transverse spin direction, whereas the quadrupole
term, proportional to $\cos 2\phi$, generates the characteristic
elongation of the density, which in the present ansatz is aligned with
the spin direction (lower panel of Fig.~\ref{fig:1}). The separate
monopole, dipole, and quadrupole layers are displayed in
Fig.~\ref{fig:2}. These patterns originate from the interference among
different helicity amplitudes and are the $x$-resolved counterparts of
the angular structures observed in the transition charge densities at
the form-factor level~\cite{Carlson:2007xd}.

Finally, the visualization presented in this section is based on a
simple multipole ansatz and illustrates the angular patterns of the
transition densities rather than their absolute magnitudes. A
quantitative description of the multipole GPDs requires dynamical
input. First estimates exist in the forward limit, where the
tensor-polarized density was computed in the mean-field
approach~\cite{Kim:2025ilc, Kim:2025ves}. Extending these
calculations to nonzero transverse momentum transfer within the
$\chi$QSM gives access to the dipole and quadrupole multipole GPDs
and is under way.

\section{Summary and conclusions}
\label{sec:5}
In the present work, we have established the multipole structure of
the $N \to \Delta$ transition at the level of the generalized parton
distributions. Evaluating the light-front Rarita--Schwinger and Dirac
spinors in the symmetric frame, we performed a multipole expansion of
the covariant matrix element in terms of the three-dimensional
spin-transition tensors and the transverse momentum transfer. The four
covariant transition GPDs were decomposed into one monopole
($M_{0}$), two dipole ($M^{V}_{1}$ and $M^{Q}_{1}$), and one
quadrupole ($M_{2}$) components in the transverse plane. Each
multipole GPD corresponds to a definite light-front helicity
amplitude, the helicity flip of which matches the rank of the
transverse multipole tensor. The explicit relations between the
multipole and covariant GPDs were also derived.

In the zero-skewness limit, the multipole GPDs define the
impact-parameter transition densities, which extend the transverse
transition charge densities from the form-factor level to the
$x$-dependent level. For longitudinally polarized states, only the
monopole component contributes, whereas for transversely polarized
states the dipole and quadrupole components distort the density in the
transverse plane. We emphasized that the transition density must be
distinguished from the diagonal densities of the nucleon and the
$\Delta$. It arises from the overlap of two distinct hadronic states
and obeys neither a positivity condition nor a number sum 
rule. The quadrupole pattern of the transition density thus
characterizes the spin-angular structure of the $N \to \Delta$
overlap rather than the intrinsic deformation of either baryon alone.

The multipole GPDs introduced in the present work provide a
model-independent framework for analyzing the partonic structure of
the $N \to \Delta$ transition, which will be probed by hard exclusive
reactions at the JLab upgrades and the EIC. A quantitative evaluation
of the multipole GPDs within the $\chi$QSM, which naturally realizes
the large-$N_{c}$ spin-flavor symmetry, is under way and will be
reported elsewhere. 

\section*{Acknowledgments}
The authors express their gratitude to C{\'e}dric Lorc{\'e} for
invaluable discussions and hospitality during their visit to the
Centre de Physique Th{\'e}orique (CPHT) at {\'E}cole polytechnique,
where part of the present work was carried out. They are also grateful
to the members of CPHT for their warm welcome and thank H.-Y. Won for
valuable discussions. JYK is grateful to C. Weiss, 
K. Semenov-Tian-Shansky, H.-D. Son, and S.-Y. Son for useful
discussions on transition densities. The present work was supported by
the National Research Foundation of Korea (NRF) grant funded by the
Korea government under Grant No.~RS-2025-00513982 (HChK) 
  and by the Future-Generation Core Researcher Joint Research Program
  of Inha University under the 4th Stage BK21 Graduate School
  Innovation Program (Project No.~78368-8) (JYK).%

\appendix
\section{Matrix elements of spin-multipole
  operators \label{app:a}}
The explicit matrix elements of the spin-multipole structures
appearing in the multipole expansion of the nonlocal vector-current
matrix element in Eq.~\eqref{eq:multipole_slashn} are listed
below. The rows correspond to $\sigma'=3/2,\,1/2,\,-1/2,\,-3/2$ from 
top to bottom, and the columns to $\sigma=1/2,\,-1/2$ from left to
right: 
\begin{subequations}
\begin{align}
Q^{33} &=  \left(\begin{array}{c c }
0 &  0 \\
\sqrt{\frac{2}{3}} &  0 \\
0 &  -\sqrt{\frac{2}{3}} \\
0 &  0 \\
\end{array} \right), \\[1ex]
Q^{3i} X^i_1 &=  \left(\begin{array}{c c }
-\frac{1}{2\sqrt{2}} e^{-i\theta} &  0 \\
0 &  \frac{1}{2} \sqrt{\frac{3}{2}} e^{-i\theta} \\
\frac{1}{2} \sqrt{\frac{3}{2}} e^{i\theta} & 0 \\
0 &  -\frac{1}{2\sqrt{2}} e^{i\theta} \\
\end{array} \right), \\[1ex]
i \epsilon^{3kj} V^{j} X^k_1 &=  \left(\begin{array}{c c }
-\frac{\sqrt{3}}{2} e^{-i\theta} &  0 \\
0 &  -\frac{1}{2}  e^{-i\theta} \\
-\frac{1}{2}  e^{i\theta} & 0 \\
0 &  -\frac{\sqrt{3}}{2} e^{i\theta} \\
\end{array} \right),\\[1ex]
Q^{ij} X^{ij}_2 &=  \left(\begin{array}{c c }
0 & -\frac{1}{\sqrt{2}} e^{-2i\theta} \\
0 &  0 \\
0 & 0 \\
\frac{1}{\sqrt{2}}e^{2i\theta}  &  0 \\
\end{array} \right).
\end{align}
\end{subequations}

\section{Light-front Rarita--Schwinger
  and Dirac spinors \label{app:b}}
In this Appendix, we collect the light-front~(LF) expressions for the
Rarita--Schwinger and Dirac spinors. The LF Rarita--Schwinger spinor
is obtained by coupling a spin-1 polarization vector to a spin-$1/2$
Dirac spinor through the Clebsch--Gordan
coefficients~\cite{Rarita:1941mf}: 
\begin{align}
u^{\alpha}(p,\sigma)=
\sum_{\lambda s}
C^{\frac32 \sigma}_{1 \lambda \frac12 s}
\epsilon^{\alpha}(p,\lambda) u(p,s).
\label{eq:RS}
\end{align}
In LF quantization, the Dirac spinor is given by~\cite{Chung:1991st,
  deAraujo:1999ugw, Rinehimer:2009yv, Lorce:2017isp} 
\begin{align}
u(p,s)=\frac{1}{\sqrt{\sqrt{2}p^+}}
(\slashed{p}+m) \slashed{\bar n} \, \chi_{s},
\label{eq:DS}
\end{align}
where $\chi_{s}=(\phi_{s},0)$ denotes the rest-frame Dirac spinor,
with $\phi_{s}$ being the two-component spinor. The dimensionless LF
vector is defined as $\bar{n}=(1,0,0,-1)/\sqrt{2}$, and $m$ denotes
the on-shell mass. The explicit representation of Eq.~\eqref{eq:DS}
reads~\cite{Lepage:1980fj, Brodsky:1997de} 
\begin{align}
u\left(p,+\frac12\right)
&=
\frac{1}{\sqrt{2\sqrt{2}p^+}}
\begin{pmatrix}
\sqrt{2}p^+ + m \\
p_R \\
\sqrt{2}p^+ - m \\
p_R
\end{pmatrix},
\\[2mm]
u\left(p,-\frac12\right)
&=
\frac{1}{\sqrt{2\sqrt{2}p^+}}
\begin{pmatrix}
-p_L \\
\sqrt{2}p^+ + m \\
p_L \\
-\sqrt{2}p^+ + m
\end{pmatrix},
\end{align}
where $p_{R,L}=p^x\pm i p^y$. The spin-1 polarization vector is given
by 
\begin{subequations}
\begin{align}
\epsilon^{\alpha} (p,\lambda =\pm 1) &=  \left( 0,
                                       \mp\frac{1}{\sqrt{2}}
                                       \frac{p_{R,L}}{p^{+}},
                                       \hat{\bm{\epsilon}}_{\perp
                                       \pm1} \right), \\ 
\epsilon^{\alpha} (p,\lambda =0) &= \left( \frac{p^{+}}{m},
                                   \frac{\bm{p}^{ 2}_{\perp} -
                                   m^{2}}{2m p^{ +}},
                                   \frac{\bm{p}_{\perp}}{m}\right). 
\end{align}
\end{subequations}
The transverse components of the rest-frame polarization vectors
$\hat{\bm{\epsilon}}_{\lambda}$ and momentum $\bm{p}$ are given by 
\begin{align}
\hat{\epsilon}^{i}_{\perp \pm 1} = \mp \sqrt{\frac{1}{2}} \left(1,\pm
  i\right), \qquad p^{i}_{\perp}= (p^x,p^y). 
\end{align}
With these conventions, Eq.~\eqref{eq:RS} gives the explicit LF
Rarita--Schwinger spinor. 

\section{Light-front spinor bilinears \label{app:c}}
In this Appendix, we present the calculation of the bilinear matrix
elements constructed from the LF Rarita--Schwinger and Dirac spinors
for each spin projection, using the conventions summarized in
Appendix~\ref{app:b}. In the symmetric frame discussed in
Sec.~\ref{sec:SF}, the Dirac bilinears can be expressed as 
\begin{subequations}
\label{eq:diracLFsym}
\begin{align}
  \bar{u}(p^{\prime},s^{\prime}) \slashed{n} \gamma_{5} u(p,s)
  &=  2  \sqrt{1-\xi^{2}}  \, \sigma^{3},
    \label{eq:diracLFsyma}
  \\[1ex] 
\bar{u}(p^{\prime},s^{\prime}) \gamma_{5} u(p,s) &=
                                                   \frac{1}{\sqrt{1-\xi^{2}}
                                                   }\left(\kappa
                                                   \sigma^{3}-
                                                   \Delta^{j}_{\perp}
                                                   \sigma^{j}\right),
                                                   \label{eq:diracLFsymb}
\end{align}
\end{subequations}
where the spin indices are suppressed, with
\begin{align}
\sigma^{i}\equiv \phi^{\dagger}_{s'}\sigma^{i}\phi_{s}.
\end{align}
The kinematic factor $\kappa$ in Eq.~\eqref{eq:diracLFsymb} is defined
as 
\begin{align}
\kappa = m_{\Delta}- m_{N} +\xi (m_{\Delta}+ m_{N}).
\end{align}

For the spin-1 polarization vector appearing in the final-state Rarita--Schwinger spinor, it is useful to record the following relations in the symmetric frame at nonzero $\xi$. We have
\begin{subequations}
\begin{align}
\epsilon^{+ *}(p',\lambda')
&= \frac{P^{+}(1-\xi)}{m_{\Delta}} \delta_{\lambda' 0}, \\[1ex]
\epsilon^{*}(p',\lambda') \cdot \Delta 
&= a_{1} \delta_{\lambda' 0}
+ a_{2}\, \hat{\bm{\epsilon}}^{*}_{\perp \lambda'} \cdot \bm{\Delta}_{\perp},
  \end{align}
  \end{subequations}
where
\begin{subequations}
\label{eq:V_con}
\begin{align}
  a_{1}
  &= \frac{P^{+}(1-\xi)}{m_{\Delta}}\Delta^{-}
- \frac{\xi (\bm{\Delta}^{2}_{\perp} - 4m^{2}_{\Delta})}{4(1-\xi)m_{\Delta}}
- \frac{\bm{\Delta}^{2}_{\perp}}{2m_{\Delta}}, \\
  a_{2}
  &= -\frac{1}{1-\xi}.
  \end{align}
  \end{subequations}

Using the notation in Eqs.~\eqref{eq:diracLFsym}--\eqref{eq:V_con},
the matrix elements of the spinor bilinears are derived as 
\begin{subequations}
\begin{align}
&\bar{u}^{\alpha}(p^{\prime},\sigma^{\prime}) \Delta_{\alpha}
                \slashed{n} \gamma_{5} u(p,\sigma) =2
                \sqrt{1-\xi^{2}}   \cr 
&\times \bigg{[}a_{1}Q^{33}  + a_2 Q^{3k} \Delta^{k}_{\perp} +
                                          \frac{a_{2}}{2}
                                          \sqrt{\frac{2}{3}} i
                                          \epsilon^{3kj} V^{j}
                                          \Delta^{k}_{\perp}\bigg{]},
  \\ 
&\bar{u}^{\alpha}(p^{\prime},\sigma^{\prime}) \Delta_{\alpha}
        \gamma_{5} u(p,\sigma) = \frac{1}{\sqrt{1-\xi^{2}}}  \cr 
&\times  \bigg{[} \kappa a_{1} Q^{33} + (\kappa a_{2}-a_1)
                Q^{3k}\Delta^{k}_{\perp} \cr 
&   +\sqrt{\frac{2}{3}}\frac{\kappa a_{2}+a_1}{2} i \epsilon^{3kj}
                                                V^{j}
                                                \Delta^{k}_{\perp}  -
                                                a_{2} Q^{lk}
                                                \Delta^{l}_{\perp}
                                                \Delta^{k}_{\perp}\bigg{]},
  \\ 
&\bar{u}^{\alpha}(p^{\prime},\sigma^{\prime}) n_{\alpha} \gamma_{5}
        u(p,\sigma) = \frac{1-\xi }{m_{\Delta}\sqrt{1-\xi^{2}}}  \cr 
&\times \bigg{[} \kappa  Q^{33} -  Q^{3k}\Delta^{k}_{\perp}   +
             \frac{1}{2} \sqrt{\frac{2}{3}} i \epsilon^{3kj} V^{j}
                 \Delta^{k}_{\perp}\bigg{]}, \\  
&\bar{u}^{\alpha}(p^{\prime},\sigma^{\prime}) n_{\alpha}  \slashed{n}
                                                    \gamma_{5}
                                                    u(p,\sigma) =
                                                    \frac{2(1-\xi)}{m_{\Delta}}
                                                    \sqrt{1-\xi^{2}}
                                                    Q^{33}. 
\end{align}
\end{subequations}

\section{Relations between covariant
  and multipole GPDs \label{app:d}}
Matching the covariant GPD parametrization to the multipole expansion
of the nonlocal vector-current matrix element, we obtain the following
relations between the multipole GPDs and the covariant GPDs: 
\begin{subequations}
\begin{align}
M_{0}&(x,\xi,t) = \frac{2\sqrt{2}}{\sqrt{3}\sqrt{1-\xi^{2}}} \cr
&\times \bigg{[} \left\{ \frac{(1-\xi^{2})}{m_{N}}a_1
          -\frac{(1-\xi)\bar{m}}{m_{N}m_{\Delta}} \kappa \right\} G_{1} \cr 
&+\left\{ \frac{\kappa a_1}{m^{2}_{N}} - \frac{\delta_{N\Delta}
        \bar{m} (1-\xi)}{m^{2}_{N} m_{\Delta}} \kappa
      + \frac{|\bm{\Delta}_{\perp}|^{2}}{2m^{2}_{N}} a_2 \right\} G_{2} \cr 
  &-\left\{ \frac{\xi \kappa a_1}{m^{2}_{N}}
    + \frac{(1-\xi)t}{2m^{2}_{N}m_{\Delta}} \kappa
    +  \frac{\xi |\bm{\Delta}_{\perp}|^{2}}{2m^{2}_{N}} a_2 \right\} G_{3} \cr
  &+\left\{ \frac{m_{N}}{m_{\Delta}} (1-\xi)(1-\xi^{2}) \right\} G_{X}
    \bigg{]}, \\
M^{Q}_{1}&(x,\xi,t) = \frac{2\sqrt{2}}{\sqrt{3}\sqrt{1-\xi^{2}}} \cr
  &\times \bigg{[} \left\{ (1-\xi^{2})a_2 +
    \frac{(1-\xi)\bar{m}}{m_{\Delta}} \right\} G_{1} \cr
  &+\left\{ \frac{\kappa a_2 -a_1}{m_{N}}
    + \frac{\delta_{N\Delta} \bar{m} (1-\xi)}{m_{N} m_{\Delta}}
    \right\} G_{2} \cr
  &-\left\{ \xi \frac{\kappa a_2 -a_1}{m_{N}}
    - \frac{(1-\xi)t}{2m_{N}m_{\Delta}} \right\} G_{3} \bigg{]},  \\
M^{V}_{1}&(x,\xi,t) = - \frac{2}{3\sqrt{1-\xi^{2}}} \cr
  &\times\bigg{[} \left\{ (1-\xi^{2})a_2
    - \frac{(1-\xi)\bar{m}}{m_{\Delta}} \right\} G_{1} \cr
  &+\left\{ \frac{\kappa a_2 +a_1}{m_{N}}
    - \frac{\delta_{N\Delta} \bar{m} (1-\xi)}{m_{N}m_{\Delta}}
    \right\} G_{2} \cr
  &-\left\{ \xi\frac{\kappa a_2 +a_1}{m_{N}}
    + \frac{(1-\xi)t}{2m_{N}m_{\Delta}} \right\} G_{3} \bigg{]},  \\
  M_{2}&(x,\xi,t) = -\frac{2\sqrt{2}}{\sqrt{3}\sqrt{1-\xi^{2}}}
         \bigg{[} a_2  G_2 - \xi a_2 G_3 \bigg{]},
\end{align}
\end{subequations}
where $\bar{m}=(m_\Delta + m_N)/2$ and $\delta_{N\Delta}= m_\Delta -
m_N$.
%
%
\bibliography{NDeltaGPDs}

@article{Rarita:1941mf,
    author = "Rarita, William and Schwinger, Julian",
    title = "{On a theory of particles with half integral spin}",
    doi = "10.1103/PhysRev.60.61",
    journal = "Phys. Rev.",
    volume = "60",
    pages = "61",
    year = "1941"
}

@article{Brodsky:1997de,
    author = "Brodsky, Stanley J. and Pauli, Hans-Christian and Pinsky, Stephen S.",
    title = "{Quantum chromodynamics and other field theories on the light cone}",
    eprint = "hep-ph/9705477",
    archivePrefix = "arXiv",
    reportNumber = "SLAC-PUB-7484, MPIH-V1-1997",
    doi = "10.1016/S0370-1573(97)00089-6",
    journal = "Phys. Rept.",
    volume = "301",
    pages = "299--486",
    year = "1998"
}

@article{Lepage:1980fj,
    author = "Lepage, G. Peter and Brodsky, Stanley J.",
    title = "{Exclusive Processes in Perturbative Quantum Chromodynamics}",
    reportNumber = "SLAC-PUB-2478",
    doi = "10.1103/PhysRevD.22.2157",
    journal = "Phys. Rev. D",
    volume = "22",
    pages = "2157",
    year = "1980"
}

@article{Chung:1991st,
    author = "Chung, P. L. and Coester, F.",
    title = "{Relativistic constituent quark model of nucleon form-factors}",
    doi = "10.1103/PhysRevD.44.229",
    journal = "Phys. Rev. D",
    volume = "44",
    pages = "229--241",
    year = "1991"
}

@article{deAraujo:1999ugw,
    author = "de Araujo, W. R. B. and Beyer, M. and Frederico, T. and Weber, H. J.",
    title = "{Feynman versus Bakamjian-Thomas in light front dynamics}",
    eprint = "hep-ph/9904307",
    archivePrefix = "arXiv",
    doi = "10.1088/0954-3899/25/8/303",
    journal = "J. Phys. G",
    volume = "25",
    pages = "1589--1592",
    year = "1999"
}

@article{Rinehimer:2009yv,
    author = "Rinehimer, Jared A. and Miller, Gerald A.",
    title = "{Connecting the Breit Frame to the Infinite Momentum Light Front Frame: How G(E) turns into F(1)}",
    eprint = "0902.4286",
    archivePrefix = "arXiv",
    primaryClass = "nucl-th",
    reportNumber = "NT@UW-09-05",
    doi = "10.1103/PhysRevC.80.015201",
    journal = "Phys. Rev. C",
    volume = "80",
    pages = "015201",
    year = "2009"
}

@article{Lorce:2017isp,
    author = "Lorc{\'e}, C{\'e}dric",
    title = "{New explicit expressions for Dirac bilinears}",
    eprint = "1705.08370",
    archivePrefix = "arXiv",
    primaryClass = "hep-ph",
    doi = "10.1103/PhysRevD.97.016005",
    journal = "Phys. Rev. D",
    volume = "97",
    number = "1",
    pages = "016005",
    year = "2018"
}

@article{Burkardt:2000za,
    author = "Burkardt, Matthias",
    title = "{Impact parameter dependent parton distributions and off forward parton distributions for zeta ---{\ensuremath{>}} 0}",
    eprint = "hep-ph/0005108",
    archivePrefix = "arXiv",
    doi = "10.1103/PhysRevD.62.071503",
    journal = "Phys. Rev. D",
    volume = "62",
    pages = "071503",
    year = "2000",
    note = "[Erratum: Phys.Rev.D 66, 119903 (2002)]"
}

@article{Burkardt:2002hr,
    author = "Burkardt, Matthias",
    title = "{Impact parameter space interpretation for generalized parton distributions}",
    eprint = "hep-ph/0207047",
    archivePrefix = "arXiv",
    doi = "10.1142/S0217751X03012370",
    journal = "Int. J. Mod. Phys. A",
    volume = "18",
    pages = "173--208",
    year = "2003"
}

@article{Carlson:2007xd,
    author = "Carlson, Carl E. and Vanderhaeghen, Marc",
    title = "{Empirical transverse charge densities in the nucleon and the nucleon-to-Delta transition}",
    eprint = "0710.0835",
    archivePrefix = "arXiv",
    primaryClass = "hep-ph",
    reportNumber = "WM-07-109, JLAB-THY-07-738",
    doi = "10.1103/PhysRevLett.100.032004",
    journal = "Phys. Rev. Lett.",
    volume = "100",
    pages = "032004",
    year = "2008"
}

@book{Varshalovich:1988ifq,
    author = "Varshalovich, D. A. and Moskalev, A. N. and Khersonskii, V. K.",
    title = "{Quantum Theory of Angular Momentum}: {Irreducible Tensors, Spherical Harmonics, Vector Coupling Coefficients, 3nj Symbols}",
    doi = "10.1142/0270",
    isbn = "978-981-4415-49-1, 978-9971-5-0107-5",
    publisher = "World Scientific Publishing Company",
    year = "1988"
}

@article{Jones:1972ky,
    author = "Jones, H. F. and Scadron, M. D.",
    title = "{Multipole gamma N Delta form-factors and resonant photoproduction and electroproduction}",
    doi = "10.1016/0003-4916(73)90476-4",
    journal = "Annals Phys.",
    volume = "81",
    pages = "1--14",
    year = "1973"
}

@article{Kim:2024hhd,
    author = "Kim, June-Young and Semenov-Tian-Shansky, Kirill M. and Won, Ho-Yeon and Son, Sangyeong and Weiss, Christian",
    title = "{Complete definition of N{\textrightarrow}{\ensuremath{\Delta}} transition generalized parton distributions}",
    eprint = "2501.00185",
    archivePrefix = "arXiv",
    primaryClass = "hep-ph",
    reportNumber = "JLAB-THY-24-4253",
    doi = "10.1103/m4vj-bkqy",
    journal = "Phys. Rev. D",
    volume = "111",
    number = "11",
    pages = "114010",
    year = "2025"
}

@article{Kim:2025ves,
    author = "Kim, June-Young",
    title = "{Tensor-polarized parton density in the  N{\textrightarrow}{\ensuremath{\Delta}} transition from the large-Nc light-cone wave function}",
    eprint = "2508.11491",
    archivePrefix = "arXiv",
    primaryClass = "hep-ph",
    reportNumber = "JLAB-THY-25-4492",
    doi = "10.1103/93ym-mbyj",
    journal = "Phys. Rev. D",
    volume = "113",
    number = "11",
    pages = "114019",
    year = "2026"
}

@article{Kim:2025ilc,
    author = "Kim, June-Young and Weiss, Christian",
    title = "{Tensor-polarized parton density in the N{\textrightarrow}{\ensuremath{\Delta}} transition}",
    eprint = "2507.18402",
    archivePrefix = "arXiv",
    primaryClass = "hep-ph",
    reportNumber = "JLAB-THY-25-4423",
    doi = "10.1016/j.physletb.2025.139924",
    journal = "Phys. Lett. B",
    volume = "870",
    pages = "139924",
    year = "2025"
}

@article{Belitsky:2005qn,
    author = "Belitsky, A. V. and Radyushkin, A. V.",
    title = "{Unraveling hadron structure with generalized parton distributions}",
    eprint = "hep-ph/0504030",
    archivePrefix = "arXiv",
    reportNumber = "JLAB-THY-04-34",
    doi = "10.1016/j.physrep.2005.06.002",
    journal = "Phys. Rept.",
    volume = "418",
    pages = "1--387",
    year = "2005"
}

@article{Diehl:2003ny,
    author = "Diehl, M.",
    title = "{Generalized parton distributions}",
    eprint = "hep-ph/0307382",
    archivePrefix = "arXiv",
    reportNumber = "DESY-THESIS-2003-018",
    doi = "10.1016/j.physrep.2003.08.002",
    journal = "Phys. Rept.",
    volume = "388",
    pages = "41--277",
    year = "2003"
}

@article{Goeke:2001tz,
    author = "Goeke, K. and Polyakov, Maxim V. and Vanderhaeghen, M.",
    title = "{Hard exclusive reactions and the structure of hadrons}",
    eprint = "hep-ph/0106012",
    archivePrefix = "arXiv",
    doi = "10.1016/S0146-6410(01)00158-2",
    journal = "Prog. Part. Nucl. Phys.",
    volume = "47",
    pages = "401--515",
    year = "2001"
}

@article{Ji:1996ek,
    author = "Ji, Xiang-Dong",
    title = "{Gauge-Invariant Decomposition of Nucleon Spin}",
    eprint = "hep-ph/9603249",
    archivePrefix = "arXiv",
    reportNumber = "MIT-CTP-2517",
    doi = "10.1103/PhysRevLett.78.610",
    journal = "Phys. Rev. Lett.",
    volume = "78",
    pages = "610--613",
    year = "1997"
}

@article{Pascalutsa:2006up,
    author = "Pascalutsa, Vladimir and Vanderhaeghen, Marc and Yang, Shin Nan",
    title = "{Electromagnetic excitation of the Delta(1232)-resonance}",
    eprint = "hep-ph/0609004",
    archivePrefix = "arXiv",
    reportNumber = "JLAB-THY-06-537",
    doi = "10.1016/j.physrep.2006.09.006",
    journal = "Phys. Rept.",
    volume = "437",
    pages = "125--232",
    year = "2007"
}

@article{PDG:2026,
    author = "Takahashi, F. and others",
    collaboration = "Particle Data Group",
    title = "{Review of particle physics}",
    journal = "Int. J. Mod. Phys. A",
    volume = "41",
    pages = "2630011",
    year = "2026"
}

@article{Muller:1994ses,
    author = "M{\"u}ller, D. and Robaschik, D. and Geyer, B. and Dittes, F. -M. and Ho{\v{r}}ej{\v{s}}i, J.",
    title = "{Wave functions, evolution equations and evolution kernels from light ray operators of QCD}",
    eprint = "hep-ph/9812448",
    archivePrefix = "arXiv",
    doi = "10.1002/prop.2190420202",
    journal = "Fortsch. Phys.",
    volume = "42",
    pages = "101--141",
    year = "1994"
}

@article{Radyushkin:1997ki,
    author = "Radyushkin, A. V.",
    title = "{Nonforward parton distributions}",
    eprint = "hep-ph/9704207",
    archivePrefix = "arXiv",
    doi = "10.1103/PhysRevD.56.5524",
    journal = "Phys. Rev. D",
    volume = "56",
    pages = "5524--5557",
    year = "1997"
}

@article{Diehl:2024bmd,
    author = "Diehl, S. and others",
    title = "{Exploring baryon resonances with transition generalized parton distributions: status and perspectives}",
    eprint = "2405.15386",
    archivePrefix = "arXiv",
    primaryClass = "hep-ph",
    doi = "10.1140/epja/s10050-025-01552-2",
    journal = "Eur. Phys. J. A",
    volume = "61",
    number = "6",
    pages = "131",
    year = "2025"
}

@article{CLAS:2023akb,
    author = "Diehl, S. and others",
    collaboration = "CLAS",
    title = "{First Measurement of Hard Exclusive $\pi^- \Delta^{++}$ Electroproduction Beam-Spin Asymmetries off the Proton}",
    eprint = "2303.11762",
    archivePrefix = "arXiv",
    primaryClass = "hep-ex",
    doi = "10.1103/PhysRevLett.131.021901",
    journal = "Phys. Rev. Lett.",
    volume = "131",
    number = "2",
    pages = "021901",
    year = "2023"
}

@article{Kroll:2022roq,
    author = "Kroll, P. and Passek-Kumeri{\v{c}}ki, K.",
    title = "{Transition GPDs and exclusive electroproduction of $\pi$-$\Delta$(1232) final states}",
    eprint = "2211.09474",
    archivePrefix = "arXiv",
    primaryClass = "hep-ph",
    doi = "10.1103/PhysRevD.107.054009",
    journal = "Phys. Rev. D",
    volume = "107",
    number = "5",
    pages = "054009",
    year = "2023"
}

@article{Semenov-Tian-Shansky:2023bsy,
    author = "Semenov-Tian-Shansky, Kirill M. and Vanderhaeghen, Marc",
    title = "{Deeply virtual Compton process $e^- N \to e^- \gamma \pi N$ to study nucleon to resonance transitions}",
    eprint = "2303.00119",
    archivePrefix = "arXiv",
    primaryClass = "hep-ph",
    doi = "10.1103/PhysRevD.108.034021",
    journal = "Phys. Rev. D",
    volume = "108",
    number = "3",
    pages = "034021",
    year = "2023"
}

@article{Kim:2019gka,
    author = "Kim, June-Young and Kim, Hyun-Chul",
    title = "{Electromagnetic transition form factors, $E2/M1$ and $C2/M1$ ratios of the baryon decuplet}",
    eprint = "2002.05980",
    archivePrefix = "arXiv",
    primaryClass = "hep-ph",
    journal = "Eur. Phys. J. C",
    volume = "80",
    number = "11",
    pages = "1087",
    year = "2020"
}

@article{AlexandrouDeformation,
    author = "Alexandrou, Constantia and Papanicolas, Costas N. and Vanderhaeghen, Marc",
    title = "{Colloquium: The shape of hadrons}",
    eprint = "1201.4511",
    archivePrefix = "arXiv",
    primaryClass = "hep-ph",
    doi = "10.1103/RevModPhys.84.1231",
    journal = "Rev. Mod. Phys.",
    volume = "84",
    pages = "1231",
    year = "2012"
}

@article{Beck:1997ew,
    author = "Beck, R. and others",
    title = "{Measurement of the E2/M1 ratio in the $N \to \Delta$ transition using the reaction $p(\vec{\gamma},p)\pi^0$}",
    doi = "10.1103/PhysRevLett.78.606",
    journal = "Phys. Rev. Lett.",
    volume = "78",
    pages = "606--609",
    year = "1997"
}

@article{Beck:1999ge,
    author = "Beck, R. and others",
    title = "{Determination of the E2/M1 ratio in the $\gamma N \to \Delta(1232)$ transition from a simultaneous measurement of $p(\vec{\gamma},p)\pi^0$ and $p(\vec{\gamma},\pi^+)n$}",
    eprint = "nucl-ex/9908017",
    archivePrefix = "arXiv",
    doi = "10.1103/PhysRevC.61.035204",
    journal = "Phys. Rev. C",
    volume = "61",
    pages = "035204",
    year = "2000"
}

@article{Blanpied:1997zz,
    author = "Blanpied, G. and others",
    collaboration = "LEGS",
    title = "{$N \to \Delta$ Transition from Simultaneous Measurements of $p(\vec{\gamma},\pi)$ and $p(\vec{\gamma},\gamma)$}",
    doi = "10.1103/PhysRevLett.79.4337",
    journal = "Phys. Rev. Lett.",
    volume = "79",
    pages = "4337--4340",
    year = "1997"
}

@article{Blanpied:2001ae,
    author = "Blanpied, G. and others",
    collaboration = "LEGS",
    title = "{$N \to \Delta$ transition and proton polarizabilities from measurements of $p(\vec{\gamma},\gamma)$, $p(\vec{\gamma},\pi^0)$, and $p(\vec{\gamma},\pi^+)$}",
    doi = "10.1103/PhysRevC.64.025203",
    journal = "Phys. Rev. C",
    volume = "64",
    pages = "025203",
    year = "2001",
    note = "[Erratum: Phys. Rev. C 64, 029902 (2001)]"
}

@article{Mertz:1999hp,
    author = "Mertz, C. and others",
    title = "{Search for quadrupole strength in the electroexcitation of the $\Delta^+(1232)$}",
    eprint = "nucl-ex/9902012",
    archivePrefix = "arXiv",
    doi = "10.1103/PhysRevLett.86.2963",
    journal = "Phys. Rev. Lett.",
    volume = "86",
    pages = "2963--2966",
    year = "2001"
}

@article{Sparveris:2004jn,
    author = "Sparveris, N. F. and others",
    collaboration = "OOPS",
    title = "{Investigation of the conjectured nucleon deformation at low momentum transfer}",
    eprint = "nucl-ex/0408003",
    archivePrefix = "arXiv",
    doi = "10.1103/PhysRevLett.94.022003",
    journal = "Phys. Rev. Lett.",
    volume = "94",
    pages = "022003",
    year = "2005"
}

@article{Pospischil:2000ad,
    author = "Pospischil, T. and others",
    collaboration = "A1",
    title = "{Measurement of the recoil polarization in the $p(\vec{e},e'\vec{p})\pi^0$ reaction at the $\Delta(1232)$ resonance}",
    eprint = "nucl-ex/0010020",
    archivePrefix = "arXiv",
    doi = "10.1103/PhysRevLett.86.2959",
    journal = "Phys. Rev. Lett.",
    volume = "86",
    pages = "2959--2962",
    year = "2001"
}

@article{Frolov:1998pw,
    author = "Frolov, V. V. and others",
    title = "{Electroproduction of the $\Delta(1232)$ resonance at high momentum transfer}",
    eprint = "hep-ex/9808024",
    archivePrefix = "arXiv",
    doi = "10.1103/PhysRevLett.82.45",
    journal = "Phys. Rev. Lett.",
    volume = "82",
    pages = "45--48",
    year = "1999"
}

@article{Joo:2001tw,
    author = "Joo, K. and others",
    collaboration = "CLAS",
    title = "{$Q^2$ dependence of quadrupole strength in the $\gamma^* p \to \Delta^+(1232) \to p \pi^0$ transition}",
    eprint = "hep-ex/0110007",
    archivePrefix = "arXiv",
    doi = "10.1103/PhysRevLett.88.122001",
    journal = "Phys. Rev. Lett.",
    volume = "88",
    pages = "122001",
    year = "2002"
}

@article{Ungaro:2006df,
    author = "Ungaro, M. and others",
    collaboration = "CLAS",
    title = "{Measurement of the $N \to \Delta^+(1232)$ transition at high momentum transfer by $\pi^0$ electroproduction}",
    eprint = "hep-ex/0606042",
    archivePrefix = "arXiv",
    doi = "10.1103/PhysRevLett.97.112003",
    journal = "Phys. Rev. Lett.",
    volume = "97",
    pages = "112003",
    year = "2006"
}

@article{Aznauryan:2009mx,
    author = "Aznauryan, I. G. and others",
    collaboration = "CLAS",
    title = "{Electroexcitation of nucleon resonances from CLAS data on single pion electroproduction}",
    eprint = "0909.2349",
    archivePrefix = "arXiv",
    primaryClass = "nucl-ex",
    doi = "10.1103/PhysRevC.80.055203",
    journal = "Phys. Rev. C",
    volume = "80",
    pages = "055203",
    year = "2009"
}

@article{Kamalov:1999hs,
    author = "Kamalov, S. S. and Yang, Shin Nan",
    title = "{Pion cloud and the $Q^2$ dependence of $\gamma^* N \leftrightarrow \Delta$ transition form factors}",
    eprint = "nucl-th/9904072",
    archivePrefix = "arXiv",
    doi = "10.1103/PhysRevLett.83.4494",
    journal = "Phys. Rev. Lett.",
    volume = "83",
    pages = "4494--4497",
    year = "1999"
}

@article{Sato:2000jf,
    author = "Sato, T. and Lee, T. -S. H.",
    title = "{Dynamical study of the $\Delta$ excitation in $N(e,e'\pi)$ reactions}",
    eprint = "nucl-th/0010025",
    archivePrefix = "arXiv",
    doi = "10.1103/PhysRevC.63.055201",
    journal = "Phys. Rev. C",
    volume = "63",
    pages = "055201",
    year = "2001"
}

@article{AbdulKhalek:2021gbh,
    author = "Abdul Khalek, R. and others",
    title = "{Science Requirements and Detector Concepts for the Electron-Ion Collider: EIC Yellow Report}",
    eprint = "2103.05419",
    archivePrefix = "arXiv",
    primaryClass = "physics.ins-det",
    doi = "10.1016/j.nuclphysa.2022.122447",
    journal = "Nucl. Phys. A",
    volume = "1026",
    pages = "122447",
    year = "2022"
}
\end{document}